\documentstyle[twocolumn,aps,epsf]{revtex}

\begin{document}

\draft
\title{Statistical Complexity of Simple 1D Spin Systems}

\author{James P. Crutchfield}
\address{Physics Department,  University of California, Berkeley,
CA 94720-7300\\
and Santa Fe Institute, 1399 Hyde Park Road, Santa Fe, NM 87501\\
Electronic Address:  chaos@gojira.berkeley.edu}
\vspace{-0.05in}
\author{David P. Feldman}
\address{Department of Physics, University of California, Davis,
CA 95616\\
Electronic Address: dfeldman@landau.ucdavis.edu}
\date{\today}
\maketitle
\begin{abstract}
We present exact results for two complementary measures of spatial
structure generated by 1D spin systems with finite-range interactions.
The first, excess entropy, measures the apparent spatial memory stored
in configurations. The second, statistical complexity, measures the
amount of memory needed to optimally predict the chain of spin values.
These statistics capture distinct properties and
are different from existing thermodynamic quantities.
\end{abstract}

\vspace{-0.07in}
\vspace{-0.3in}


Thermodynamic entropy, a measure of disorder, is a
familiar quantity that is well-understood in almost all
statistical mechanical contexts. It's notable, though, that
complementary and similarly general measures of structure and
pattern are largely missing from current theory and are certainly
less well-developed. To date, ``structure'' has been handled on a
case by case basis. Order parameters and structure functions, for
example, are typically invented to capture the significant features
in a specific phenomenon. There is no generally accepted approach
to answering relatively simple questions, such as, How much temporal
memory is used by a process to produce a given level of disorder?

In the following we adapt two measures of structure, the excess
entropy $E$ and the statistical complexity $C_\mu$, to analyze the
spatial configurations generated by simple spin systems. These measures
of structure are not problem-specific---they may be applied to any 
statistical mechanical system. We give exact results for $E$ and
$C_\mu$ as a function of temperature, external field, and coupling
strength for one-dimensional finite-range systems. Our results
show that $E$ and $C_\mu$ are different from measures of disorder,
such as thermodynamic entropy and temperature; rather $E$ and $C_\mu$
quantify significant aspects of information storage and computation
embedded in spatial configurations.

In our analysis we introduce purely information theoretic
coordinates---a plot of $E$ and $C_\mu$ vs. spatial entropy density
$h_\mu$ -- known as the complexity-entropy diagram. The benefit of this
view is that it is explicitly independent of system parameters and so
allows very different systems to be compared directly in terms of
their intrinsic information processing. In past work the
complexity-entropy diagram was analyzed for a class of processes in
which the set of allowed configurations changed as a function of a
system control parameter \cite{INFERRING}. For the systems considered
here, the variation in $E$ and $C_\mu$ is driven instead by the
``thermalization'' of the configuration distribution.


Consider a one-dimensional chain of spin variables
$  \,\stackrel{\leftrightarrow}{s} \, = \ldots \, s_{-2} \, s_{-1} \, s_0 \, s_1 \, \ldots $ 
where $s_i$ range over a finite set
${\cal A}$.
Divide the chain into two semi-infinite halves by choosing a site $i$
as the dividing point. Denote the left half by
$  \,\stackrel{\leftarrow}{s_i} \, \, \equiv \, \ldots \, s_{i-3} \, s_{i-2} \, s_{i-1} \, s_i \,$
and the right half by
$\, \stackrel{\rightarrow}{s}_i \equiv s_{i+1} \, s_{i+2} \, s_{i+3} \, \ldots \;$ .
Let ${\rm Pr}\,(s_i)$ denote the probability that the
$i^{{\underline{th}}}$ variable takes on the particular value $s_i$
and ${\rm Pr}\, ( s_i, s_{i+1} , \ldots , s_{i+L})$ 
the joint probability over blocks of $L$ consecutive spins.
Assuming spatial translation symmetry,
${\rm Pr}\,( s_i, \ldots , s_{i+L} ) = {\rm Pr}\,(s_1, \ldots , s_L )$.

Given such a distribution one measures the average uncertainty of
observing a given $L$-spin block $s^L$ by the Shannon entropy
\cite{COVER}
\begin{eqnarray}
   \; \; \; H(L)&  \equiv &  - \sum_{s_1 \in \cal A}\, \ldots \, 
	\sum_{s_L \in \cal A} \nonumber \\ 
	& &   {\rm Pr}\,(s_1, \ldots , s_L )
	\log_2 {\rm Pr}\,(s_1 , \ldots , s_L )   \; .  \; \; 
\label{shannon}
\end{eqnarray}
The spatial density of Shannon entropy of the spin configurations is
defined by $h_\mu \equiv \lim_{L \rightarrow \infty} L^{-1} H(L)$.
$h_\mu$ measures the irreducible randomness in the spatial 
configurations. For physical systems it is, up to a multiplicative constant,
equivalent to thermodynamic entropy density. It is
also equivalent to the average of the configurations'
Kolmogorov-Chaitin complexity. As such, $h_\mu$ measures the average
length (per spin) of the minimal universal Turing machine program
required to produce a typical configuration \cite{COVER,VITANYI}. 

The entropy density is a property of the system as a whole; only
in special cases will the isolated-spin uncertainty $H(1)$ be equal
to $h_\mu$. It is natural to ask, therefore, how random the chain
of spins appears when finite-length spin blocks are considered.
This is given by $h_\mu (L) \equiv \,H(L) - H(L-1)$, the incremental
increase in uncertainty in going from $(L-1)$-blocks to $L$-blocks.
$h_\mu (L)$ overestimates the entropy density $h_\mu$ by an amount
$h_\mu (L) - h_\mu$ that indicates how much more random the finite $L$
blocks appear than the infinite configuration $ \,\stackrel{\leftrightarrow}{s} \,$. In other words,
this excess randomness tells us how much additional information must
be gained about the configurations in order to reveal the actual
per-spin uncertainty $h_\mu$. Summing up the overestimates one obtains
the total excess entropy \cite{EXCESS}
\begin{equation}
	E \equiv \sum_{L=1}^\infty ( h_\mu(L) - h_\mu ) \; .
\label{excess}
\end{equation}
Informally, $E$ is the amount (in bits), above and beyond $h_\mu$, of
{\it apparent} randomness that is eventually ``explained'' by
considering increasingly longer spin-blocks. This follows from noting
that $E$ may be expressed as the mutual information $I$ \cite{COVER}   
between the two semi-infinite halves of a configuration;
$E = I(  \,\stackrel{\leftarrow}{s} \, ; \, \stackrel{\rightarrow}{s})$. That is, $E$ measures how much information
one half of the spin chain carries about the other. In this restricted
sense $E$ measures the spin system's apparent spatial memory. If the
configurations are perfectly random or periodic with period 1, then
$E$ vanishes. Excess entropy is nonzero between the two extremes of
ideal randomness and trivial predictability.


Another, related, approach to spatial structure begins by asking a
different question, How much memory is needed to optimally predict
configurations? Restated, we are asking to model the system in such
a way that the observed configurations can be statistically reproduced.
To address this, we must determine the effective states
of the process; how much of the left configuration must be remembered
to optimally predict the right? The answer to these questions leads us
to define the statistical complexity $C_\mu$ \cite{INFERRING}.

Consider the probability distribution of all possible right halves
$\, \stackrel{\rightarrow}{s}$ conditioned on a particular left half, $ \,\stackrel{\leftarrow}{s_i} \,$ at site $i$: 
${\rm Pr}(\, \stackrel{\rightarrow}{s}\, |  \,\stackrel{\leftarrow}{s_i} \,) $. These conditional probabilities allow
one to optimally predict configurations. We now use this form of
conditional probabilities to define an equivalence relation $\sim\,$
on the space of all left halves; the induced equivalence classes are
subsets of the space of all allowed $ \,\stackrel{\leftarrow}{s_i} \,$. We say that two
configurations at different lattice sites are equivalent if and only
if they give rise to an identical conditional distribution of
right-half configurations. Formally, we define the relation $\sim\,$ by
\begin{equation}
     \,\stackrel{\leftarrow}{s_i} \, \sim  \,\stackrel{\leftarrow}{s_{j}} \, \; {\rm iff} \;\, \rm Pr (\, \stackrel{\rightarrow}{s}\, |  \,\stackrel{\leftarrow}{s_i} \,) =
	\rm Pr (\, \stackrel{\rightarrow}{s}\, |  \,\stackrel{\leftarrow}{s_{j}} \, ) \; \;\; \forall \, \stackrel{\rightarrow}{s} \; .
\label{equiv}
\end{equation}
The equivalence classes induced by this relation are called
{\em causal states} and denoted ${\cal S}_i$. Two $ \,\stackrel{\leftarrow}{s} \,$ belong to 
same causal state if, as measured by the probability distribution of
subsequent spins conditioned on having seen that particular left-half
configuration, they give rise to exactly the same degree of certainty
about the configurations that follow to the right.  

Once the set $\lbrace {\cal S}_i \rbrace$ of causal states has been
identified, we can inductively obtain the probability
${\rm Pr}({\cal S}_i)$ of finding the chain in the
$i^{{\underline{th}}}$ causal state by observing many configurations.
Similarly, we can obtain transition probabilities $T$ between states.
The set $\lbrace {\cal S}_i \rbrace$ together with the dynamic $T$
constitute a model---referred to as an $\epsilon$-machine
\cite{INFERRING}---of the original infinite configurations.  

To predict, as one scans from left to right, the successive spins in
a configuration with an $\epsilon$-machine, one must track in which
causal state the process is. Thus, the informational size of the
distribution over causal states gives the minimum amount of memory 
needed to optimally predict the right-half configurations. This
quantity is the statistical complexity
\begin{equation}
    C_\mu \equiv - \sum_{ \lbrace {\cal S}_i \rbrace }
	{\rm Pr}({\cal S}_i) \log_2 {\rm Pr} ({\cal S}_i) \; .
\end{equation}

The excess entropy sets a lower bound on the statistical complexity:
$E \leq C_\mu$ \cite{ONSET}. That is, the memory needed to perform
optimal prediction of the right-half configurations cannot be lower
than the mutual information between left and right halves themselves.   
This relationship reflects the fact that the set of causal states is
not in one-to-one correspondence with $L$-block or even $\infty$-length
configurations. In the most general setting, the causal states are a
reconstruction of the hidden, effective states of the process.

Note that for both $C_\mu$ and $E$ no memory is expended trying to
account for the randomness or, in this case, for thermal fluctuations
present in the system. Thus, these measures of structural complexity
depart markedly from Kolmogorov-Chaitin complexity which demands a
deterministic accounting for the value of every spin in a configuration.
As noted above, the per-spin Kolmogorov-Chaitin complexity is $h_\mu$
\cite{COVER,VITANYI}. Finally, note that $C_\mu$ and $E$ follow
directly from the configuration distribution; their
calculation doesn't require knowledge of the Hamiltonian.


As is well known, the partition function for any one-dimensional,
discrete spin system with finite range interactions can be
expressed in terms of the transfer matrix $V$ \cite{transfer}.  
Using $V$, we have calculated exact expressions for $C_\mu$ and $E$
for such systems. In the following let $u^{\cal R}$ ($ u^{\cal L} $)
denote the normalized right (left) eigenvector corresponding to $V$'s
largest eigenvalue $\lambda$.

The first step is to determine the causal states. Consider an Ising
system with nearest neighbor (nn) interactions. The nn interactions
and the fact that a configuration's probability is determined by the
temperature and its energy means that only the rightmost spin in the
left half influences the probability distribution of the spins in the
right half. Thus, the possible causal states are in a one-to-one
correspondence with the different values of a single spin.  (This
indicates how this class of spin systems is a severely restricted
subset of $\epsilon$-machines.) This observation determines
an upper bound for a spin 1/2 nn system:
$C_\mu \leq \log_2 2 =1 $.

To complete the determination of the causal states we must verify that
conditioning on different spin values leads to {\em different}
distributions for $\, \stackrel{\rightarrow}{s}$; otherwise they fall into the same
equivalence class and there would be only one causal state.
This distinction is given by eq.~(\ref{equiv}) which,
in terms of the transfer matrix $V$, reads
\begin{equation}
    ( u^{\cal R}_i )^{-1} V_{ik} \, \neq \, 
	 ( u^{\cal R}_j )^{-1} V_{jk} \; \; \forall\, i \neq j \; . 
\label{question}
\end{equation}
If eq.\ (\ref{question}) is satisfied, then
\begin{equation}
	C_\mu = - u^{\cal L}_k u^{\cal R}_k\log_2 ( u^{\cal L}_k u^{\cal R}_k ) \; .\
\label{Cmu}
\end{equation}
(In eq.~(\ref{Cmu}) and the following, a summation over repeated
indices is implied.) For a nn system, eq.\ (\ref{Cmu}) is 
equivalent to saying that $C_\mu = H(1)$, the entropy associated with the
value of one spin.
By determining an expression for $H(L)$, one sees that $h_\mu$ is given by
$ 
 h_\mu = \log \lambda - {\lambda}^{-1} u^{\cal R}_i  u^{\cal L}_k  
        V_{ki} \log [ \, V_{ki} \, ] \; ,
\label{h2}
$
and that $E$ is given by
\begin{eqnarray}
   E & = & - \log \lambda \, + \, \frac{1}{\lambda}  
	u^{\cal R}_i  u^{\cal L}_k 
        V_{ki}\, \log[ \, V_{ki} \, ] \, -  \\ \nonumber
        & & \; \; \; u^{\cal L}_k   u^{\cal R}_k  \, 
	\log [ \,  u^{\cal R}_k  u^{\cal L}_k \, ] \; ,
\label{e2}
\end{eqnarray}
Note that these results prove an explicit version of the
inequality between $E$ and $C_\mu$ mentioned above;
namely,
\begin{equation}
C_\mu \, = \, E +  h_\mu \; ,
\label{relationship}
\end{equation}
again assuming that  eq.\ (\ref{question}) is
satisfied \cite{GENERALIZE}.


Let us illustrate the content of eq.\ (\ref{question}) by considering
a special case, a spin 1/2 paramagnet (PM), where there are no couplings
between spins. Since there are no correlations between spins, 
$E$ vanishes. The probability distribution of the right-half
configuration is independent of the left-half configuration. Thus,
there is a single, unique distribution ${\rm Pr}(\, \stackrel{\rightarrow}{s}\, |  \,\stackrel{\leftarrow}{s} \,) $
and eq.\ (\ref{question}) is not satisfied. The PM has only
one causal state and so $C_\mu = 0$ for all temperatures.
This example shows how the process of determining
causal states ensures statistical complexity measures structure and
not randomness.

\vspace{-.1in}
\begin{figure}
\begin{center}
\leavevmode
\hbox{%
\epsfxsize=3.0in
\epsfysize=1.80in
\epsffile{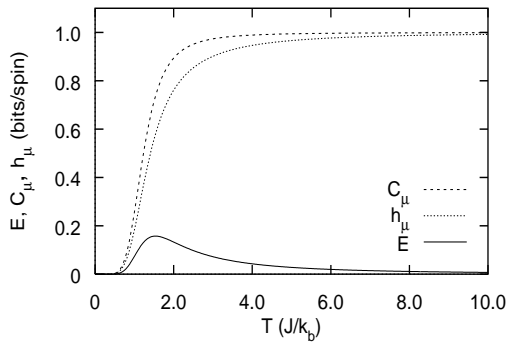}}
\end{center}
\caption{$C_\mu$, $E$, and $h_\mu$ as a function of $\rm T$ for the
nn spin 1/2 ferromagnet. $B$ was held at $0.30$ and $J = 1$.}
\label{FIG1}
\end{figure}

Now consider the spin 1/2, nearest neighbor Ising system with
Hamiltonian
\begin{equation}
	{\cal{H}}  =  -J \sum_{i} s_i s_{i+1}  - B \sum_i s_i  \, ,
\label{HamH}
\end{equation}
where, as usual, $J$ is a parameter determining the strength of
coupling between spins, $B$ represents an external field, and
$s_i \in \{+1, -1\}$.

For all temperatures except zero and infinity eq.~(\ref{question})
is satisfied and the causal states are in a one-to-one correspondence
with the values of a single spin. At ${\rm T} = \infty$ the system is
identical to a paramagnet and $C_\mu$ and $E$ both vanish. At $T=0$
the system is frozen in its spatially periodic ground state;
$E = C_\mu = \log_2 P = 0$, where $P (= 1)$ is the period of the
spatial pattern.

Using eqs.\ (\ref{Cmu}) and (\ref{e2}), fig. 1 plots $C_\mu$ and $E$
as a function of temperature $\rm T$. The coupling is ferromagnetic
($J=1$) and there is a non-zero external field ($B=0.3$). As expected,
the entropy density is a monotonic increasing function of temperature.
Somewhat less expectedly (cf.~ref.~\cite{INFERRING}), the statistical
complexity also increases monotonically (until ${\rm T} = \infty$). The
excess entropy $E$ vanishes gradually in the high and low temperature
limits.

\vspace{-.1in}
\begin{figure}
\begin{center}
\leavevmode
\hbox{%
\epsfxsize=3.0in
\epsfysize=1.80in
\epsffile{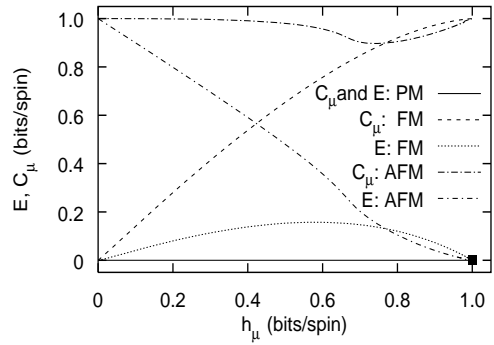}}
\end{center}
\caption{The complexity-entropy diagram for a ferromagnet (FM), an
anti-ferromagnet (AFM) and a paramagnet (PM): $C_\mu$ and $E$ plotted
parametrically against $h_\mu$. For a given $J$, $B$ was held
constant---$B = 0.30$ (FM) and $B = 1.8$ (AFM)---as $\rm T$ was
varied.  All systems have $C_\mu = 0$ when $h_\mu = 1$;
this is denoted by the square token.}
\label{FIG2}
\end{figure}
Figure 2 presents the complexity-entropy diagram for a ferromagnet (FM),
an anti-ferromagnet (AFM), and a paramagnet (PM): $C_\mu$ and $E$
plotted parametrically as a function of $h_\mu$. The diagram gives
direct access to the information processing properties of the systems
independent of control parameters (i.e. $B$, $J$, and $\rm T$).

For the ferromagnet, $E$ is seen to have a maximum in a region between
total randomness ($h_\mu = 1$) and complete order ($h_\mu = 0$).  
At low temperatures (and, hence, low $h_\mu$) most of the spins
line up with the magnetic field.  At high temperatures, thermal noise
dominates and the configurations are quite random. In both regimes
one half of a configuration contains very little information about the
other half. For low $h_\mu$, the spins are fixed and so there is no
information to share; for high $h_\mu$, there is much information at
each site, but it is uncorrelated with all other sites. Thus, the
excess entropy is small in these temperature regimes. In between the
extremes, however, $E$ has a unique maximum at the temperature where
spin coupling strength balances the thermalization. The result is a
maximum in the system's spatial memory.

For an AFM, the high temperature behavior is similar;
thermal fluctuations destroy all correlations and  $E$ vanishes.
The low $\rm T$ behavior is different; the ground state of
the AFM consists of alternating up and down spins.  The spatial
configurations thus store one bit of information about whether the odd
or even sites are up. As can be seen in fig.~2,
$E \rightarrow 1$ as $h_\mu \rightarrow 0$.

For different couplings and field strengths a range of $E$ vs.\ $h_\mu$
relationships can be realized. $E$ either shows a single maximum or
decreases monotonically.  It is always the case, though, that $E$ is
bounded from above by $1 - h_\mu$, which follows immediately if
$C_\mu$ is set equal to its maximum value, 1, in
eq.\ (\ref{relationship}).

Given that $C_\mu$ was introduced as a measure of structure, it is
perhaps surprising that it behaves so differently from $E$. As $h_\mu$
increases, one might expect $C_\mu$ to reach a maximum, as does $E$,
and then decrease as the increasing thermalizing merges causal
states that were distinct at lower temperatures. In fact, $C_\mu$
increases monotonically with $h_\mu$. To understand this, recall that
the causal states are the same for all $\rm T$ between zero and
infinity. For the nn spin 1/2 Ising model, the number of 
causal states remains fixed at two.  What {\em does} change as
${\rm T}$ is varied are the causal state probabilities. For the
FM, as the temperature rises the distribution
${\rm Pr} ( {\cal S}_i )$ becomes more uniform,  
and $C_\mu$ grows.  This growth continues until ${\rm T}$ becomes
infinite, since only there do the two causal states collapse into one,
at which point $C_\mu$ vanishes.

For the AFM the situation is a little different. At ${\rm T} = 0$
there are two causal states corresponding to the two spatial phases
of the alternating up-down pattern.  The probability of these causal
states is uniform; hence we see a low temperature statistical complexity
of 1.  At high (but finite) temperatures, the thermal
fluctuations dominate; the anti-ferromagnetic order is lost, but the
distribution over causal states is still relatively uniform so the
statistical complexity remains large.  (As with the FM, at
${\rm T} = \infty$
the two causal states merge and $C_\mu$ jumps to zero.) Between these
extremes there is a region where the influence of the external field
dominates, biasing the configurations. This is reflected in a bias
in the causal state probabilities and $C_\mu$ dips below 1 as seen
in fig.~2.

The tendency for $C_\mu$ to remain large for large values of $h_\mu$
is due to a more general effect, which follows from
eq.\ (\ref{relationship}): $ C_\mu = E + h_\mu$. The memory needed 
to model a process depends not only on the internal memory of the
process, as measured by $E$, but also on its randomness, as
measured by $h_\mu$. It is important to note, however, that $C_\mu$
is driven up by thermalization {\em not} because the model attempts
to account for random spins in the configuration. Rather, $C_\mu$
rises with $h_\mu$ because $\Pr ( {\cal S}_i )$ becomes 
more uniform as the temperature increases.


We have discussed three complementary statistics that as a whole capture
the information processing capabilities embedded in spin systems. This
framework has been applied previously to the symbolic dynamics of
continuous-state dynamical systems \cite{INFERRING}.
The work presented here is the first 
exploration of thermal systems with these tools. In the dynamical 
systems studied, the statistical complexity varied as a function of
$h_\mu$ mainly due to changes in topological constraints on
configurations. This led to changes in the number of causal states and
in their connectivity. As a result, $C_\mu$ has a unique maximum at
some $h_\mu < 1$.  (cf.~Fig ??, ref. \cite{INFERRING}.)
In sharp contrast, the thermal systems examined
here have the same number of causal states for all temperatures except
zero and infinity. For all ${\rm T} \neq 0$ thermal fluctuations
are present: all configurations are possible and the
connectivity of the causal states remains the same. This contrast
points out a possibly useful distinction between deterministic
and stochastic systems---a distinction that is lost by comparing
these two different types of process solely in terms of $h_\mu$.

These features and other work to be reported indicate that $E$ and
$C_\mu$ capture properties that are different from existing
thermodynamic quantities. Comparing the PM, FM, and AFM in terms of
specific heat, for example, doesn't reveal the distinctions seen in
fig.~2. This issue---along with analyses of 2D Ising systems, spin
glasses, and recurrent neural networks---will be discussed elsewhere.
For these higher dimensional systems, there are a number of ways
to define $E$ and $C_\mu$.  One approach is to consider infinite
strips of spins as a single, infinite-dimensional spin.  This 
method involves a natural extension of the techniques developed here,
yet we feel this might not faithfully capture the higher
dimensional structure present.  Another approach is to adapt the
cellular automata-theoretic formalism presented in ref.~\cite{CA}.  
We shall examine both of these approaches in a future work.

We thank Richard T. Scalettar for many helpful comments and suggestions.
This work was supported at UC Berkeley by ONR grant N00014-95-1-0524
and AFOSR grant 91-0293 and at the Santa Fe Institute by NASA-Ames
contract NCC2-840 and ONR grant N00014-95-1-0975.



\vspace{-0.2in}
\begin{references}

\vspace{-0.65in}
\bibitem{INFERRING} J.P.~Crutchfield and K.~Young,
Phys. Rev. Lett. {\bf 63}, 105 (1989); J.P.~Crutchfield, 
Physica D {\bf 75}, 11 (1994).

\bibitem{COVER} T.M.~Cover and J.A.~Thomas.
\newblock {\em Elements of Information Theory} 
(John Wiley \& Sons, Inc., 1991).

\bibitem{VITANYI} M.~Li and P.M.B.~Vitanyi.  {\em An Introduction to
Kolmogorov Complexity and its Applications} (Springer-Verlag, 1993).

\bibitem{EXCESS} J.P.~Crutchfield and N.H.~Packard, Physica D {\bf 7},
201 (1983); P.~Sz\'{e}pfalusy and G.~Gy\"{o}rgyi, Phys. Rev. A {\bf 33},
2852 (1986); cf. ``stored information'',
R.~Shaw, {\em The Dripping Faucet as a Model Chaotic System}
(Aerial Press, 1984); cf. ``effective measure complexity'',
P.~Grassberger, Int. J. Theo. Phys. {\bf 25}, 907 (1986);
K.~Lindgren and M.~Nordahl, Complex Systems {\bf 2}, 409 (1988);
and cf. ``complexity'', W.~Li, Complex Systems {\bf 5}, 381 (1991). 

\bibitem{ONSET} J.P.~Crutchfield and K.~Young, in
{\em Complexity, Entropy and the Physics of Information},
edited by W.~H. Zurek, (Addison-Wesley, 1990) 223.

\bibitem{transfer} 
H.A.~Kramers and G.H.~Wannier, Phys.~Rev.~ {\bf 145}, 251 (1941);
J.F.~Dobson, J.~Math.~ Phys. {\bf 10} 1, 40  (1969).

\bibitem{GENERALIZE} For $r^{th}$ nn interactions, the causal states
are the values of $r$-spin blocks. Although the dimensionality of
$V$ increases, our results remain unchanged. However, if $V$
adds on the effects of $m$ spins at a time, then our expression
for $h_\mu$ must be divided by $m$.
Details of our calculations will be presented elsewhere.

\bibitem{CA} J.P.~Crutchfield and J.E.~Hanson.  Physica {\bf D}.
In Press.


\end{references}
\end{document}